\documentclass[prb,aps,preprint]{revtex4}
\usepackage{amsmath}
\usepackage{amsfonts}
\usepackage{amssymb}
\usepackage{graphicx}
\usepackage{epsfig}
\usepackage{bm}
\usepackage{multirow}
\usepackage{rotating}

\begin{document}
%\date{}

\title{LaSrVMoO$_6$: A case study for $A$-site covalency-driven local cationic order in double perovskites }

\author{Somnath Jana,$^{1}$ %
Vijay Singh,$^{2}$ %
Abhishek Nag,$^{5}$ %
Giuliana Aquilanti,$^{3}$%
Indra Dasgupta,$^{1,2}$ %
Carlo Meneghini,$^{4}$ %
Sugata Ray$^{1,5,\star}$}

\affiliation{$^1$Centre for Advanced Materials, Indian Association for the Cultivation of Science, Jadavpur, Kolkata 700 032, India\\ %
$^2$Department of Solid State Physics, Indian Association for the Cultivation of Science, Jadavpur, Kolkata 700 032, India\\ %
$^3$Sincrotrone Trieste S.C.p.A., s.s. 14,  km 163.5 34149 Basovizza, Trieste, Italy\\%
$^4$Dipartimento di Fisica E. Amaldi, Universit\'{a} di Roma Tre, Via della vasca navale, 84 I-00146 Roma, Italy\\ %
$^5$Department of Materials Science, Indian Association for the Cultivation of Science, Jadavpur, Kolkata 700 032, India}

\begin{abstract}
An unusual atomic scale chemical fluctuation in LaSrVMoO$_6$, in terms of narrow patches of La,V and Sr,Mo-rich phases, has been probed in detail to understand the origin of such a chemical state. Exhaustive tuning of the equilibrium synthesis parameters showed that the extent of phase separation can never be melted down below an unit cell dimension making it impossible to achieve the conventional $B$-site ordered structure, which establishes that the observed `inhomogeneous' patch-like structure with minimum dimension of few angstroms is a reality in LaSrVMoO$_6$. Therefore, another type of local chemical order, hitherto unknown in double perovskites, gets introduced here. X-ray diffraction, electron microscopy elemental mapping, magnetic, and various spectroscopic studies have been carried out on samples, synthesized under different conditions. These experimental results in conjunction with {\it ab-initio} electronic structure calculation revealed that it is the energy stability, gained by typical La-O covalency as in LaVO$_3$, that leads to the preferential La,V and Sr,Mo ionic proximity, and the consequent patchy structure.
\end{abstract}

\maketitle

PACS number(s): 75.47.Lx, 61.66.Fn, 71.15.Mb, 72.25.Ba

%\newpage

\section{Introduction}
%In the recent times an increasing interest is found in the multicationic systems in order to achieve more exiciting functionalities.
The $B$ site of a $AB$O$_3$ perovskite is commonly doped to achieve exciting functionalities. In such doped compounds, the dopant ($B^\prime$) and the parent ($B$) atoms  at commensurate doping often organize themselves to create interesting ordering patterns. Among many such orders, an ideal double perovskite of general formula $A_2BB^\prime$O$_6$ is realized only by half doping a single perovskite and by alternately arranging the $B,B^\prime$ cations in a rock-salt-type order. However, the degree of this $B,B^\prime$ ordering is known to vary under different conditions~\cite{ssc} and complex patterns are created; in some cases the true local order may even go invisible to long range structural probes.~\cite{xafs_prl} The complexity increases many fold for $AA'BB'$O$_6$ double perovskites with an additional cation, where quite a few different ionic ordering patterns become possible.~\cite{woodward_JMC} However, unlike all previous descriptions, LaSrVMoO$_6$ (LSVMO) presented a completely distinct cationic arrangement, as was revealed during the investigation of its debatable half-metallic antiferromagnetic (HMAFM) state.~\cite{somnath_PRB} Initially, this compound was suggested to be a half-metallic antiferromagnet by considering antiparallel alignment of alternately occupied V$^{3+}$ (3$d^2$) and Mo$^{4+}$ (4$d^2$) spins in a rock-salt type double perovskite structure. A couple of reports claimed this system to be the first experimentally realized HMAFM,~\cite{Uhera, Gotoh} while {\it ab-initio} band structure calculation results based on the experimental crystal structure opined otherwise,~\cite{park2,Wu_JCC} confirming half-metallicity but predicting robust ferrimagnetic ground state for this compound. Therefore, quite contradictory depictions existed till our experiments~\cite{somnath_PRB} revealed the fact that {\it the ground state of this compound is neither half-metallic nor antiferromagnetic}. Most importantly, our study showed that the local chemical order in this material is very different~\cite{somnath_PRB} from the usually perceived long range rock-salt type chemical order. Instead, a narrow scale chemical fluctuation, driven by an unusual affinity of the V(Mo) ions towards La(Sr) ions,~\cite{somnath_PRB} was found to be active in the material, and it is this unanticipated local chemical fluctuation that gives rise to all the conflicting descriptions. This result dictates that in general any special affinity between the cations ({\it e.g.} between $A$ \& $B$ and/or $A'$ \& $B'$ in $AA'BB'$O$_6$), that might be present in a multi-cation system, needs to be considered while probing the structure and properties of the material.

However, before looking for the reason of such phase fluctuations, it is necessary to check whether the observed local order~\cite{somnath_PRB} is truly the ground state configuration of LaSrVMoO$_6$ or not. As the site ordering of cations in double perovskites is known to depend critically on synthesis conditions, it might very well be possible that the observed structure is only an artefact coming due to incompetent synthesis protocols, while a microscopically homogeneous structure could actually be realized only if the very optimal synthesis conditions are applied. It is to be noted that the earlier studied sample,~\cite{somnath_PRB} synthesized following the reported synthesis parameters,~\cite{Uhera,Gotoh} appeared pure and single phase when probed by long range diffraction techniques, while the local structural study revealed a significantly different scenario.
Therefore, in this paper different attempts have been made to check, if suitably tuning the synthesis parameters, it is possible to determine a protocol that could improve the homogeneity of the system even at local scale.
More importantly, there is a possibility that the ideally proposed HMAFM behavior in this compound{\cite{Uhera,Gotoh} could actually be realized in the $B$-site ordered double perovskite structure, if it could be synthesized, and for this the foremost task would be to get rid of the local chemical fluctuation.
%Therefore, at first we have attempted to optimize the synthesis of this material by systematically tuning the parameters and carefully looking for an improvement in homogeneity.
%It was observed that the most perceptible changes occur when the synthesis atmosphere, especially the H$_2$ content is varied. In this paper, we report detailed structural, electronic and magnetic studies from three samples, %synthesized under three methodically altered H$_2$/Ar environments.
But, the present study established that the most 'homogeneous' sample is indeed the one that we have already reported earlier.~\cite{somnath_PRB} Further, our theoretical calculation showed why this structure with chemical fluctuation has to be the most stable configuration for LaSrVMoO$_6$; where the driving force is the preference of La ions to form strong covalent linkages with the O ions that are linked to V ions, over the ones connected to Mo. The distinction arises due to the more ionic nature of V ions compared to Mo ions, finally resulting in La,V-rich, and the consequent Sr,Mo-rich regions.
\section{Experimental and theoretical methods}
All the samples reported in this paper were synthesized by the conventional solid state routes. In any solid state synthesis the tunable parameters are the synthesis temperature, atmosphere, and cooling rate. Typically double perovskites are synthesized in a restricted atmosphere of H$_2$/Ar, where the relative amount of H$_2$ in the reaction atmosphere largely affects phase purity as well as physical properties of the materials. Therefore, a couple of samples is made by varying the H$_2$ content from the previously used value, although only by a small amount. We have also varied the synthesis temperature and post-synthesis cooling rate independently {\it i.e.} every time keeping all other parameters unchanged. In this paper, we discuss six different LSVMO samples, each synthesized with carefully planned and precise set of synthesis parameters (see Table I).
%, namely LSVMO-1 (ST = 1300 \textcelsius, 6\% H$_2$/Ar, CR = 5K/min), LSVMO-2 ( ST = 1300 \textcelsius, 4.5\% H$_2$/Ar, CR = 5K/min), and LSVMO-3 (ST = 1300 \textcelsius, 9\% H$_2$/Ar, CR = 5K/min), LSVMO-4 (ST = 1400 \textcelsius, 6\% H$_2$/Ar, CR = 5K/min), LSVMO-5 ( ST = 1300 \textcelsius, 6\% H$_2$/Ar, CR \~ 100K/min), and LSVMO-6 (ST = 1300 \textcelsius, 6\% H$_2$/Ar, CR = 0.5K/min).
The crystallographic structures of the samples were checked by XRD using a Bruker AXS: D8 Advance x-ray diffractometer, while the magnetic measurements were carried out in a Quantum Design SQUID magnetometer. The XRD data were analyzed using Rietveld technique and were refined using the FullProf program.~\cite{fullprof} Detailed elemental mapping using transmission electron microscopy (TEM) were performed in a JEOL2010 microscope.
EXAFS measurement at Mo K-edge were performed at the XAFS beamline of the Elettra synchrotron in Basovizza, Trieste (Italy) using a double crystal Si(311) monochromator.~\cite{xafs} The energy was calibrated by setting the first inflection point of the absorption edge of the metallic molibdenum to 20000 eV and 3 to 5 spetra with acquisition time of 5 s/pt were collected and averaged per each sample to improve the statistical noise.
The photoelectron spectroscopic (XPS) measurements were carried out in an Omicron electron spectrometer, equipped with EA125 analyzer and Mg $K_{\alpha}$ x-ray source.

The theoretical calculations were done in the framework of density-functional theory (DFT) using the generalized gradient approximation (GGA) as implemented in the Vienna ab-initio simulation package (VASP).{\cite{vasp1,vasp2} However, as it was difficult to computationally deal with the real structure with random chemical fluctuations, a {\it hypothetical superlattice} structure with periodic La,V and Sr,Mo stripes was constructed where the local atomic coordination was conformed with the experimental XAFS data.~\cite{somnath_PRB} The electron ion interaction in the core and valence part are treated within the projector augmented wave (PAW) method~\cite{vaspgga} along with the plane wave basis with an energy cut-off of 500~eV. Atomic positions were relaxed to minimize the Hellmann-Feynman forces on each atom with a tolerance value of 10$^{-2}$~ ev/{\AA}. Brillouin-zone integration were performed in a k-mesh of (4$\times$8$\times$4) and (2$\times$4$\times$2) for 8 formula unit and 16 formula unit supercells, respectively. We have analysed the chemical bonding by computing the crystal orbital Hamiltonian population(COHP) as implemented in the Stuttgart tight binding linear muffin-tin orbital (TB-LMTO) code.~\cite{lmto} The COHP provides information regarding the specific pairs of atoms that participates in the bonding and the range of such interaction.~\cite{cohp}
\section {Results and Discussions}
The XRD patterns from LSVMO-1 to LSVMO-6 are shown in Fig. 1. The best fit refinement curves are laid (red lines) over the experimental data (open circles). In the insets to Fig. 1(a) - 1(f), expanded views of the most intense peak of the XRD spectrum (at around 2$\theta$~=~32$^{\circ}$) are shown. It is to be noted that our earlier work already discussed sample LSVMO-1 in detail, which appears single phase in diffraction experiments.~\cite{somnath_PRB} Interestingly, all other samples (LSVMO-2 - LSVMO-6) indicate presence of multiple crystallographic phases even in the diffraction experiments. Firstly, as soon as the H$_2$ content is decreased (LSVMO-2) or increased (LSVMO-3) even by a very small amount with respect to the LSVMO-1 case, the system is found to be divided into more than one distinct crystallographic phases (inset to Fig. 1(b) and (c)). Similar situations are also encountered in case of LSVMO-4, LSVMO-5, and LSVMO-6 samples, albeit to a lesser extent. Evidently, LSVMO-1 is the only sample exhibiting a single symmetric peak (inset to Fig. 1(a)) that could be refined satisfactorily within a homogeneous orthorhombic structure,~\cite{somnath_PRB} while none other samples could be refined considering only one unique crystal structure. For LSVMO-2 to LSVMO-6 samples, satisfactory refinement of the diffraction patterns could be achieved only after considering existence of multiple phases having different crystallographic parameters. The goodness factors and and other refined parameters for the different phases are listed in Table-II. Careful analysis of the refinement results reveal that in every case a cubic phase with negligible orthorhombic distortion (the lower 2$\theta$ peaks in the refined spectra seen in the insets to panels (b) - (f)) coexists with phase(s) possessing significant orthorhombic distortion(s). Considering the fact that SrMoO$_3$ is cubic~\cite{srmoo3} and LaVO$_3$ possesses strong orthorhombic distortions,~\cite{lavo3} these observations in LSVMO-2 - LSVMO-6 appear only to be a natural propagation of the situation that already exists in LSVMO-1, where Sr,Mo-rich regions coexist with La,V-rich regions within very narrow spatial dimensions of few {\AA}.~\cite{somnath_PRB} However, it appears that unlike LSVMO-1, the `cubic' and `orthorhombic' regions are much more spatially extended in LSVMO-2 - LSVMO-6. In LSVMO-1, as the two phases are packed within the dimension of an unit cell, the system adopts a single, average crystal description, while in case of other samples, such regions extend spatially where maintaining unique crystallographic characters of the individual phases becomes possible. As a result, for LSVMO-2 to LSVMO-6 samples distinctly different lattice structures and parameters become visible in the diffraction experiments. It should be noted that all the multiple crystallographic phases observed in these samples could be basically categorized into two crystallographic groups, corresponding to cubic SrMoO$_3$ and orthorhombic LaVO$_3$ like phases, while only their relative concentrations differed from sample to sample.

The notion of the spatially extended phase separation in other samples compared to LSVMO-1 was more clearly probed by comparative elemental mapping of LSVMO-1 and LSVMO-2 in a high resolution TEM. The experiments were carried out with a beam of 0.7 nm diameter scanned over an area of 120$\times$160 nm$^2$ for each sample. We have optimized the data acquisition time from the signal to noise ratio of the characteristic x-rays, emitted from the elements La, Sr, V, Mo and O. The results are summarized in Fig.2(a), (b). In order to remove the fictitious contributions arising from the uneven surface topography, the intensities of La, Sr, V and Mo are normalized by oxygen intensity as this intensity is expected to be correlated only to the surface topography and unrelated to any kind of phase segregation of La,V-rich or Sr,Mo-rich regions. A qualitative glance into the four elemental maps each from LSVMO-1 and LSVMO-2, reveals that in LSVMO-1 the existence of all the elements within the given area are more or less equally probable, while there are gross inhomogeneity in LSVMO-2 (the densities of La and V are much higher than the same of Sr and Mo, in the shown area). Now, for a more quantitative description, we have divided the scanned regions into 12 small cells each having an area 40$\times$40 nm$^2$. The integrated intensities of each elements from each of these cells are plotted in the lower part of both the panels using the column bars. Now, almost in every cell, certain fluctuation in intensities along with distinct correlations between the intensities of La(Sr) and V(Mo), is observed even in case of LSVMO-1 sample, which essentially indicates presence of small La,V-rich (see cell no. 2, 3, 5 and 6) or Sr,Mo-rich (see cell no. 8, 11, 12) areas in the sample. However in case of LSVMO-2, the whole area under investigation exhibits large intensity for La and V with respect to Sr and Mo, confirming much larger spatial extension of the chemical inhomogeneity compared to that in LSVMO-1. This result strongly affirms the understanding of basic chemical nature of the samples as acquired from the XRD studies.

%The notion of larger area phase separation in LSVMO-2 and LSVMO-3 was further confirmed by energy dispersive x-ray spectroscopy (EDS) experiments in a transmission electron microscope (TEM). These experiments were carried out on all the three samples using a 5 nm wide electron beam and the results are summarized in Fig. 1(d)-(f). For a homogeneous LaSrVMoO$_6$ sample, the ratio of the atomic percentages of La and Sr should be unity at any given area, which should also be the case between V and Mo. In Fig. 1(d)-(f), the La/Sr ratio (black circles) and V/Mo ratio (red up triangles) are plotted, as collected from different areas of the samples. The data from LSVMO-1 (Fig. 1(d)) reveals that this sample is indeed nearly homogeneous within the spatial resolution of the experiment. However, both LSVMO-2 and LSVMO-3 deviate from this ideal scenario significantly (Fig. 1(e) and (f), respectively). In these cases, it was possible to identify regions with largely different La and Sr contents as well as regions with widely dissimilar V and Mo contents. However, it was reasserting that any deviation from ideal La/Sr content in a single region of the sample almost identically mapped the corresponding fluctuations in the V/Mo content, showing that La deficient areas (La/Sr~$<$~1) invariably have V deficiency and {\it vice versa}. Obviously, this observation strongly supports our prediction that similar to LSVMO-1, LSVMO-2 and LSVMO-3 also contain La,V-rich and Sr,Mo-rich phases but with significantly larger spatial extension.
In Fig. 3, the field-cooled (FC) and zero-field-cooled (ZFC) magnetization data from all the samples are shown. While LSVMO-1 and LSVMO-6 exhibit a single peak like feature at around 125 K indicating a possible magnetic transition,~\cite{somnath_PRB} other samples seem to possess multiple magnetic transitions distributed over a temperature range below 200 K. As all the samples, except LSVMO-1, have extended volumes of La,V-rich orthorhombic phase(s) and Sr,Mo-rich nearly cubic phase(s), it is likely that different areas will have different compositions and volumes having different magnetic interaction strengths, which may give rise to a group of different transition temperatures. Therefore, the observation of strong magnetic metastability in nearly all the samples is very much consistent with the perception of spatially extended phase separation in them.

Next, we have carried out local Mo $K$-edge XAFS experiments (see Fig. 4) on three of the five samples (detailed experiments on LSVMO-1 was already presented in Ref. 4). %~\cite{footnote}
Table III presents the results of the XAFS analysis: the refinement include the Mo-O nearest neighbour contribution along with next neighbours shells including the Mo-La/Sr and Mo-O-\emph{M} (\emph{M}=Mo/V), which are related to the local chemical order features.
%Due to the reduced k-range of the L-edge data and the correlation among the parameters, the incertitude on the structural parameters is quite high (in particular concerning the coordination numbers).
From the values reported in the table III,  it is evident that all the samples (LSVMO-2, LSVMO-3 and LSVMO-6) depict large affinity of Mo towards another Mo (Mo-O-Mo configurations) as well as Sr (Mo-Sr neighbours). It is apparent that samples with larger spatially extended Sr,Mo and La,V-rich phases, would show certain enhancement in the Mo-Mo and Mo-Sr coordinations compared to LSVMO-1, which is confirmed from the present set of measurements. Therefore, our present XAFS experiments strongly endorse the experimental finding that no possible synthesis protocol succeeds to destroy the preferential ionic proximities (between La,V and Sr,Mo), and therefore, formation of La,V-rich and Sr,Mo-rich patches is indeed a reality in LaSrVMoO$_6$.

%Similarly, $d. c.$ electrical resistivity measurements show (see inset to Fig. 2) that absolute resistivity is substantially smaller for LSVMO-2 and LSVMO-3 samples, compared to LSVMO-1. If the electrical conduction in this system is dominated by the SrMoO$_3$ like strongly metallic percolation channel(s), the mechanism is expected to be stronger in LSVMO-2 and LSVMO-3 compared to LSVMO-1, which is exactly what has been observed in the experiments. Therefore, all these experiments establish that the spatial extension and relative concentration of the La,V-rich and Sr,Mo-rich regions could be varied readily by just manipulating the H$_2$ contents around an optimal value (6\% H$_2$/Ar in this case) and the temperature protocols, while it appears impossible to completely destroy such preferential ionic proximities (between La,V and Sr,Mo) and create a conventional $B$-site ordered/disordered double perovskite structure.

Following this, we have also carried out detailed x-ray photoelectron spectroscopic (XPS) studies on the three samples in order to probe the effect of phase separation on the electronic structures of these compounds. The experimental results are summarized in Fig. 5. In Fig. 5(a), the height normalized O 1$s$ core level spectra from the three samples are shown. While LSVMO-1 exhibits a clean singlet spectrum, both LSVMO-2 and LSVMO-3 show clear shoulders at higher binding energy (highlighted by the striped region around 533 eV), indicating more than one type of chemical environments in these compounds. Similarly, higher binding energy features become also visible both in the Sr 3$d$ and Mo 3$d$ spectra (striped areas in Figs. 5(b) and (c)) of LSVMO-2 and LSVMO-3. It should be pointed out here that the Mo 3$d$ spectrum from a complex molybdate system is often dominated by surface oxidized $d$ components, which is hard to remove even after rigorous {\it in-situ} surface cleaning. Also, the unusually complex Mo 3$d$ spectra is a fingerprint of the prevalent covalent effects, often present in the metallic Mo-based double perovskites,~\cite{prl} which is observed here as well. Finally, the set of valence band spectra turned out to be no different, with clear higher binding energy features around 12 eV for LSVMO-2 and LSVMO-3 (striped region in Fig. 5(d)), compared to that from LSVMO-1. However, another interesting point to note is the unusually large density of states near the Fermi energy ($E_F$) in all these compounds, which clearly show presence of large amount of charge carriers that brings about the large metallicity observed in these polycrystalline samples.
%Moreover, the intensity of the DOS feature near $E_F$ ($<$2~eV binding energy) also consistently maps the resistivity behaviors, with LSVMO-1 being the most resistive member and LSVMO-2 exhibiting the least insulating nature.
Therefore, the XPS experiments reveal the existence of electronically different environments in different proportions in both LSVMO-2 and LSVMO-3, while one of them certainly agrees well with the LSVMO-1 electronic structure. Considering also the results of XRD analysis, one may therefore conclude that an additional phase with large orthorhombic distortion appears as soon as the synthesis environment shifts from the optimal condition by a small amount along any direction, which probably is electrically nonconducting, giving rise to extra features in XPS spectra, invariably shifted towards higher binding energies. Overall, XPS experiments also provided a consistent scenario.

Next, we have attempted to check the validity of these experimental findings through {\it ab-initio} electronic structure calculations. Firstly, a set of four non-spin polarized calculations were carried out on a supercell containing eight formula units of orthorhombic LaSrVMoO$_6$ with very different La, Sr, V, and Mo atomic distributions within the supercell. These model structures are shown in Fig. 6(a)-(d). In (a) and (b), the conventional $B$-site ordered and $B$-site disordered structures are shown and in both the cases a random distribution of $A$ and $A^{\prime}$ ions are maintained. However, in (c), a pair of V and Mo-rich channels stacked along c and running parallel to the b axis are created where no correlation with the corresponding La or Sr occupancies are allowed. Finally in structure (d), La,V-rich and Sr,Mo-rich {\it superlattice} structure, which is the closest approximation to the experimentally obtained structure, is shown.~\cite{somnath_PRB} For the sake of unambiguous comparison, the calculations on these four structures were carried out by keeping the k-points and the energy cut-off parameters identical. The stabilization energies of these four relaxed structures are presented in Fig. 6. From these energy values, it becomes immediately evident that structure (d), being nearly identical to the experimental structure with clear correlation between $A$ ($A^{\prime}$) and $B$ ($B^{\prime}$)-site cations, is by far the most stable structure. Therefore, a new type of possible ionic order in $AA^{\prime}BB^{\prime}$O$_6$ double perovskites, driven by spectacular affinities between a pair of $A, B$ and/or $A^{\prime}, B^{\prime}$ atoms, is introduced here.

Next, to check the ease of spatial expansion of the two different phases, we constructed a larger supercell with sixteen formula units but with two different morphologies. In one case, only the previously described superlattice structure is maintained (Fig. 6(e)), while in the other case (Fig. 6(f)) both the La,V-rich and the Sr,Mo-rich phases have been expanded along a-axis, in order to mimic the largely phase separated sample environments. It is to be noted that there is no difference between the structures shown in Fig. 6(d) and (e) and therefore, they possess the same stabilization energy. However, the energy difference between Fig. 6(e) and 6(f) turns out to be abysmally small (1.88 meV/f.u.), which basically falls within the error bar of the calculation. This again explains why it should be extremely easy to switch between LSVMO-1 (structure (e)) and other samples discussed in this paper (structure (f)), which in reality could be achieved by just changing the synthesis environment to a small extent.

Here, one could put forward an argument that the `phase separated' structures are stabilized simply because of the desire to create local LaVO$_3$ and SrMoO$_3$ like structural environments or in other words, the following chemical reaction,
\begin{center}
LaVO$_3$~+~SrMoO$_3$~$\leftrightharpoons$~LaSrVMoO$_6$~~~(1)
\end{center}
is merely inclined towards the left hand side of the reaction and never goes on to form the homogeneous LaSrVMoO$_6$ phase. However, such an explanation would fail to explain the observed energy degeneracy between the structures shown in Fig. 6(e) and (f). Because the individual LaVO$_3$ and SrMoO$_3$ structures are expected to be largely relaxed in structure (f), compared to the strongly strained structure (e), and therefore, following the above mentioned proposition, structure (f) is expected to be energetically much stable, which is in clear contradiction with our results. Hence the energy stability must stem from atomic considerations which should remain nearly unaltered between structures (e) and (f), making them energetically degenerate.
Consequently, we have carefully analyzed the chemical bonding by plotting the charge densities and COHP of all the structures along different lattice planes. This analysis revealed that all the structures gain extra energy by allowing the La ions to be covalently linked with one of the neighboring oxygens, while the Sr ions prefer to remain perfectly ionic.

In Fig. 7, we show charge density plots for a couple of lattice planes from structures (c) and (d) (defined in Fig. 6). In panel (a), we show the (002) plane charge density from structure (d). Expectedly, the V-O bond is largely covalent with the charges being nearly equally shared between vanadium and oxygen ions. Interestingly, contrary to the notion of a purely ionic La$^{3+}$ charge state, one observes a rather significant electron sharing between every La and a neighboring oxygen. Such $A$-O covalency in $AB$O$_3$ perovskite structure is known where a substantial stability is gained by asymmetrically shifting the smaller $A$ cation having extended $d$-orbital within an unit cell, thereby strengthening the $A$-O covalency and bringing forth structural distortions.~\cite{A-O_PRB} Specifically for LaVO$_3$, it has been shown that weak GdFeO$_3$-type distortion exists which is fundamentally driven by the energy gain due to La-O covalency effects.~\cite{A-O_PRB} In the present case, we find exactly the same feature dominating. On the contrary, the charge density in Sr-Mo-O plane (Fig. 7(b)) exhibits purely ionic nature of Sr. The scenario remains similar in case of structure (c), with the only exception that in this case the La and Sr ions can be randomly placed in both the V-O-V (Fig. 7(c)) and Mo-O-Mo (Fig. 7(d)) planes. Again, La is found to be involved in covalent linkages with oxygens in both the planes, while Sr prefers to be ionic everywhere. Therefore, by comparison, the only difference between the two structures is the presence of a large number of covalent linkages between the La ions with neighboring oxygen ions that are also parts of Mo-O-Mo connections in structure (c), and this must be energetically expensive so that the structure (c) becomes significantly unstable with respect to the ground state structure (d) (Fig. 6). It is expected that in a La-Mo-O plane, stronger La-O covalency would certainly compromise the Mo-O covalency and these two effects would work against each other to converge to the stable ground state. We shall now substantiate our argument by analysing the COHP plots which provide an energy resolved visualization of the chemical bonding. In COHP the density of states is weighted by the Hamiltonian matrix elements where the off-site COHP represents the covalent contribution to bands. In Fig. 7 (e, f) we display the off-site COHP and the energy integrated COHP (ICOHP) per bond for the nearest neighbor V-O, Mo-O, La-O interactions for structures (d) and (c), respectively. The bonding contribution for which the system undergoes a lowering in energy is indicated by negative COHP and antibonding contribution that adds to the energy is represented by positive COHP. From the COHP plots in Fig. 7 we find that strongest covalency is between Mo and O for both the structures. This covalency is substantially stronger in comparison to V-O covalency as revealed by the integrated COHP at the Fermi level Mo-O (-3.36eV), V-O (-1.92eV) for structure d and Mo-O (-2.63eV) ,V-O (-2.06eV) for structure c. Interestingly the nearest neighbor La-O covalency are not only appreciable but identical for both the structures (ICOHP values are (-0.87eV), (-0.77eV) for structure (d) and structure (c), respectively). Now, this can be quite convincingly argued that weakening the stronger Mo(4$d$)-O(2$p$) covalent bond would be far more energetically expensive than disturbing the V(3$d$)-O(2$p$) covalent bond and as a result presence of La-O-Mo covalent linkages would be energetically unfavorable. It is possibly this consideration which governs the unusual La,V affinity in this compound, while the Sr,Mo clustering is only a consequence of this affinity. Within this picture, it can also be understood why the structures (e) and (f) may be energetically degenerate because in both the cases nearly all the La ions find a neighboring suitable oxygen from a V-O-V linkage with which stable covalent linkages could be established and stability could be attained.

\section{Conclusions}
In summary, we show that the strong La-O covalency acts as a stabilizing factor in LaSrVMoO$_6$, while the oxygen involved in this covalent linkage prefers to be a part of a V-O-V linkage instead of a Mo-O-Mo linkage. This phenomena creates a preferential La,V clustering and as a result the unprecedent local La,V-rich and Sr,Mo-rich phase fluctuation is realized in this compound, which is extremely robust and could not be disrupted by any simplistic equilibrium synthesis methods. Probably only certain nonequilibrium growth techniques can alter the energetics and create a conventional homogeneous double perovskite structure. Overall, our results clearly establish that the local distribution of cations in any multi-cation system could be a very complicated competition between many factors that needs to be carefully addressed in order to understand such systems.

\section{Acknowledgements}
We thank S. Acharya and D. D. Sarma and their DST project SR/S5/NM-47/2005 for making the photoemission studies possible. SJ and AN thank CSIR, India for fellowship. SR thanks DST-RFBR and DST Fast Track, India for financial support. ID thanks DST India for support.

\newpage

\begin{table}
\caption{Synthesis details of polycrystalline LSVMO samples. For each sample, the variable parameter is noted in bold.}
\begin{tabular}{cccc}
\hline
Sample & Synthesis  & Atomsphere & Cooling rate  \\
 & $T$($^{\circ}$C) &  H$_2$/Ar(\%) &  ($^{\circ}$C/min) \\
\hline
\hline
LSVMO-1 & 1300  & {\bf 6} & 5  \\
%\hline
LSVMO-2 & 1300  & {\bf 4.5} & 5  \\
%\hline
LSVMO-3 & 1300  & {\bf 9} & 5  \\
%\hline
LSVMO-4 & {\bf 1400}  & 6 & 5  \\
%\hline
LSVMO-5 & 1300  & 6 & {\bf 100}  \\
%\hline
LSVMO-6 & 1300  & 6 & {\bf 0.5} \\
\hline
\end{tabular}
\end{table}

\begin{figure}
\begin{center}
%\centering
%\resizebox{7cm}{!}
%{\includegraphics*[60pt,50pt][450pt,780pt]{fig1.eps}} \\
%\vspace{-1 in}
\caption{(color online) (a)-(f) X-ray powder diffraction from the six samples at 300 K showing the observed data (open circles), calculated pattern (continuous line) and difference pattern. Insets show contributions from different crystallographic phases.}
\end{center}
\end{figure}

\begin{sidewaystable}
\caption{All the samples are refined with minimum number of phases to achieve reasonable $Rwp$ and $\chi^2$. Orthorhombic $Pnma$ space group is taken for all the phases, while Sr,Mo-rich phases are appeared to be nearly cubic after the refinement. The occupation numbers of La (Sr) and V (Mo) are constrained to be equal during the refinement, while even otherwise they appeared very close.}
\vspace*{0.5cm}
\begin{tabular}{|l|ccc|c|ccccc|}
\hline
Sample & \hspace*{0.3cm}$Rwp$ & \hspace*{0.3cm} $Rexp$ & \hspace*{0.3cm} $\chi^2$ \hspace*{0.3cm}& Phase compositions & a (\AA) & b (\AA) & c (\AA) & v (\AA$^3)$ & Fraction (\%) \\
\hline
\hline
LSVMO-1 \hspace*{0.2cm} & 8.49 & 5.02 & 2.86 & LaSrVMoO$_6$ & \hspace*{0.2cm}5.575 & \hspace*{0.2cm}7.865 & \hspace*{0.2cm}5.589 & \hspace*{0.2cm}245.050 & 100 \\
\hline
\multirow{2}{*}{LSVMO-2} \hspace*{0.2cm}& \multirow{2}{*}{7.79} & \multirow{2}{*}{3.99} & \multirow{2}{*}{3.81} & \hspace*{0.15cm}La$_{1.73}$Sr$_{0.27}$V$_{1.73}$Mo$_{0.27}$O$_6$ \hspace*{0.15cm}& \hspace*{0.2cm}5.551 & \hspace*{0.2cm}7.917 & \hspace*{0.2cm}5.601 & \hspace*{0.2cm}242.707 & 29.8 \\
        &&&& La$_{0.51}$Sr$_{1.49}$V$_{0.51}$Mo$_{1.49}$O$_6$ & \hspace*{0.2cm}5.580 & \hspace*{0.2cm}7.852 & \hspace*{0.2cm}5.568 & \hspace*{0.2cm}247.453 & 70.2  \\
\hline
\multirow{3}{*}{LSVMO-3} \hspace*{0.2cm}& \multirow{3}{*}{11.5}&\multirow{3}{*}{5.83}&\multirow{3}{*}{3.90}&La$_{1.03}$Sr$_{0.97}$V$_{1.03}$Mo$_{0.97}$O$_6$ & \hspace*{0.2cm}5.567 & \hspace*{0.2cm}7.879 & \hspace*{0.2cm}5.590 & \hspace*{0.2cm}245.180 & 40 \\
        &&&& La$_{0.34}$Sr$_{1.66}$V$_{0.34}$Mo$_{1.66}$O$_6$ & \hspace*{0.2cm}5.590 & \hspace*{0.2cm}7.935 & \hspace*{0.2cm}5.613 & \hspace*{0.2cm}248.991 & 38  \\
        &&&& La$_{1.74}$Sr$_{0.26}$V$_{1.74}$Mo$_{0.26}$O$_6$ & \hspace*{0.2cm}5.547 & \hspace*{0.2cm}7.859 & \hspace*{0.2cm}5.561 & \hspace*{0.2cm}242.409 & 22  \\
\hline
\multirow{2}{*}{LSVMO-4} \hspace*{0.2cm}&\multirow{2}{*}{10.8}&\multirow{2}{*}{4.71}&\multirow{2}{*}{5.25}& La$_{0.98}$Sr$_{1.02}$V$_{0.98}$Mo$_{1.02}$O$_6$ & \hspace*{0.2cm}5.567 & \hspace*{0.2cm}7.879 & \hspace*{0.2cm}5.592 & \hspace*{0.2cm}245.232 & 84.9  \\
        &&&& La$_{0.88}$Sr$_{1.12}$V$_{0.88}$Mo$_{1.12}$O$_6$ & \hspace*{0.2cm}5.593 & \hspace*{0.2cm}7.907 & \hspace*{0.2cm}5.556 & \hspace*{0.2cm}245.728 & 15.1  \\
\hline
\multirow{3}{*}{LSVMO-5} \hspace*{0.2cm}& \multirow{3}{*}{9.7}&\multirow{3}{*}{4.66}&\multirow{3}{*}{4.33}&LaSrVMoO$_6$ & \hspace*{0.2cm}5.567 & \hspace*{0.2cm}7.879 & \hspace*{0.2cm}5.588 & \hspace*{0.2cm}245.065 & 47.9 \\
        &&&& La$_{0.59}$Sr$_{1.41}$V$_{0.59}$Mo$_{1.41}$O$_6$ & \hspace*{0.2cm}5.592 & \hspace*{0.2cm}7.904 & \hspace*{0.2cm}5.616 & \hspace*{0.2cm}248.196 & 34.4  \\
        &&&& La$_{1.75}$Sr$_{0.25}$V$_{1.75}$Mo$_{0.25}$O$_6$ & \hspace*{0.2cm}5.552 & \hspace*{0.2cm}7.866 & \hspace*{0.2cm}5.555 & \hspace*{0.2cm}242.603 & 17.7 \\
\hline
\multirow{3}{*}{LSVMO-6} \hspace*{0.2cm}& \multirow{3}{*}{10.6}&\multirow{3}{*}{4.46}&\multirow{3}{*}{5.65}&La$_{1.03}$Sr$_{0.97}$V$_{1.03}$Mo$_{0.97}$O$_6$ & \hspace*{0.2cm}5.569 & \hspace*{0.2cm}7.880 & \hspace*{0.2cm}5.589 & \hspace*{0.2cm}245.261 & 42.6 \\
        &&&& La$_{0.49}$Sr$_{1.51}$V$_{0.49}$Mo$_{1.51}$O$_6$ & \hspace*{0.2cm}5.588 & \hspace*{0.2cm}7.903 & \hspace*{0.2cm}5.614 & \hspace*{0.2cm}247.934 & 40.8  \\
        &&&& La$_{1.8}$Sr$_{0.2}$V$_{1.8}$Mo$_{0.2}$O$_6$ & \hspace*{0.2cm}5.543 & \hspace*{0.2cm}7.863 & \hspace*{0.2cm}5.556 & \hspace*{0.2cm}242.156 & 16.6  \\
\hline
\end{tabular}
\end{sidewaystable}
\begin{figure}
\begin{center}
%\centering
%\resizebox{7cm}{!}
%{\includegraphics*[20pt,160pt][315pt,752pt]{fig2.eps}} \\
%\vspace{-1 in}
\caption{(color online) (a) Elemental mapping from LSVMO-1 (upper panel) and (b) LSVMO-2 (lower panel) samples. The total area under investigation has been devided into 12 equal area cells, shown by the overlaid lines. Integrated intensities of La, Sr, V, Mo from each of the small cells are plotted as histograms below the corresponding panel after topographic correction. Numbers of the cells are marked in the top left La sub-panel to correlate intensity bars to the corresponding cell.}
\end{center}
\end{figure}

\begin{figure}
\begin{center}
%\centering
\resizebox{10cm}{!}
{\includegraphics*[15pt,100pt][584pt,734pt]{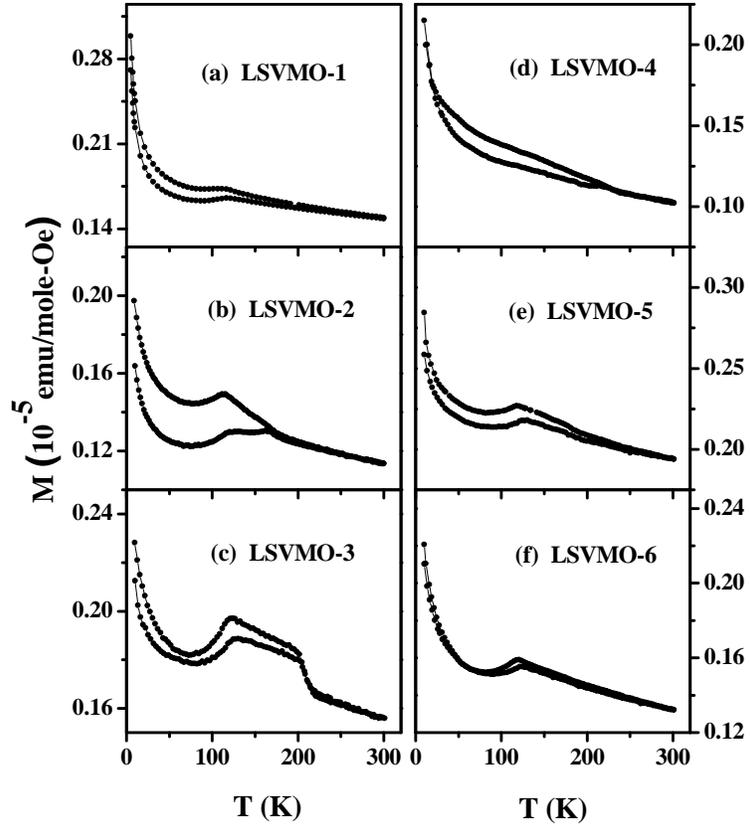}} \\
%\vspace{-1 in}
\caption{(color online) (a)-(f) ZFC (lower curve) and FC (upper curve) magnetization vs temperature data collected from six samples.}
\end{center}
\end{figure}

\begin{figure}
\begin{center}
%\centering
\resizebox{10cm}{!}
{\includegraphics*[57pt,51pt][389pt,267pt]{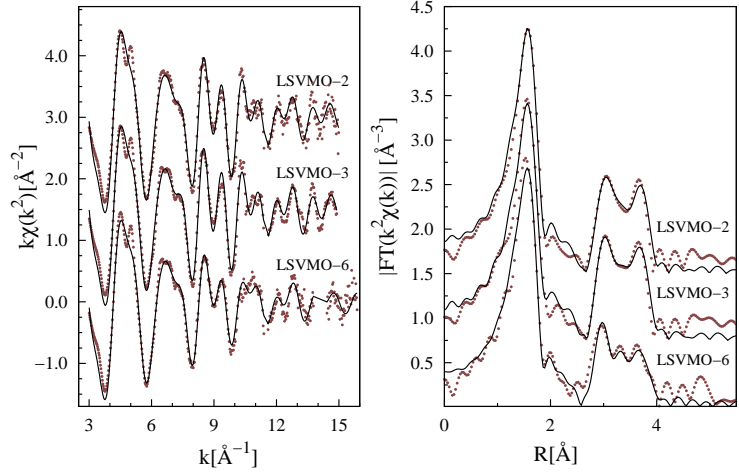}} \\
%\vspace{-1 in}
\caption{(color online) left panel: Mo $K$-edge $k^2$-weighted experimental XAFS data (points) $k\chi(k^2)$, and best fit (full line) of LSVMO-2, LSVMO-3 and LSVMO-6 samples (shifted for clarity). The right panel shows the modulus of the Fourier transform of experimental $k^2\chi(k)$ (points) and best fit (full lines).}
\end{center}
\end{figure}
\begin{sidewaystable}
\caption{Local structure parameters as obtained from the analysis of Mo $K$-edge XAFS data. In order to reduce the correlation among the parameters constraints among the parameters were applied, namely: indicating with \emph{x} the fraction of MoLa pairs, it is N$_{\mathrm {MoLa}}=8x$ and
N$_{\mathrm {MoSr}}=8(1-x)$; indicating with \emph{y} the fraction of MoOV configurations, it is N$_{\mathrm {MoOV}}=6y$ and
N$_{\mathrm {MoOMo}}=6(1-y)$. The single (SS) and multiple scattering (MS) contributions to the MoOMo (MoOV) configurations were constrained to the same $\sigma^2$. Finally the number of Mo-O bonds around 2\AA was fixed to 6. The fixed or constrained values are labeled by `*'.}
\vspace{0.3cm}
\begin{tabular}{|c|c|c|c|c|c|c|}
\hline
 & \multicolumn{3}{|c|}{LSVMO-2}& \multicolumn{3}{|c|}{LSVMO-3} \\
\hline
Shell   & N & R(\AA) & $\sigma^2 (\times 10^2$\AA$^2$)& N & R(\AA) & $\sigma^2 (\times 10^2$\AA$^2$)\\
\hline
MoO          & 6*     & 2.02(1)   & 0.38(4) & 6*     & 2.02(1)  & 0.43(4)  \\
MoLa         & 2.0(3) & 3.48(2)   & 0.29(4) & 2.1(2) & 3.48(3)   & 0.28(4) \\
MoSr         & 6.0*   & 3.49(2)   & 0.52(5) & 5.9*   & 3.49(3)   & 0.45(4) \\
MoOV(SS+MS)  & 1.6(4) & 3.99(2)   & 0.74(8) & 1.5(2) & 4.00(2)   & 0.75(8) \\
MoOMo(SS+MS) & 4.4*   & 3.98(2)   & 0.87(7) & 4.5*   & 3.98(2)   & 0.90(7) \\
\hline
\hline
& \multicolumn{3}{|c|}{LSVMO-6} & \multicolumn{3}{|c|}{LSVMO-1 (from Ref. 4)}\\
\hline
Shell   & N & R(\AA) & $\sigma^2 (\times 10^2$\AA$^2$)& N & R(\AA) & $\sigma^2 (\times 10^2$\AA$^2$)\\
\hline
MoO          & 6*     & 2.01(1)   & 0.41(4) & 6      & 1.998(5)  & 0.23(3) \\
MoLa         & 1.7(2) & 3.46(3)   & 0.32(4) & 2.2    & 3.47(2)   & 0.34(5) \\
MoSr         & 6.3*   & 3.47(3)   & 0.55(5) & 5.8(4)  & 3.46(2)   & 0.43(5) \\
MoOV(SS+MS)  & 1.2(2) & 3.99(2)   & 0.99(8) & 2.4    & 3.98(2)   & 0.69(6) \\
MoOMo(SS+MS) & 4.8*   & 3.98(2)   & 0.82(7) & 3.6(3)  & 3.97(2)   & 0.48(6) \\
\hline
\end{tabular}
\end{sidewaystable}

\begin{figure}
\begin{center}
%\centering
\resizebox{10cm}{!}
{\includegraphics*[73pt,63pt][492pt,742pt]{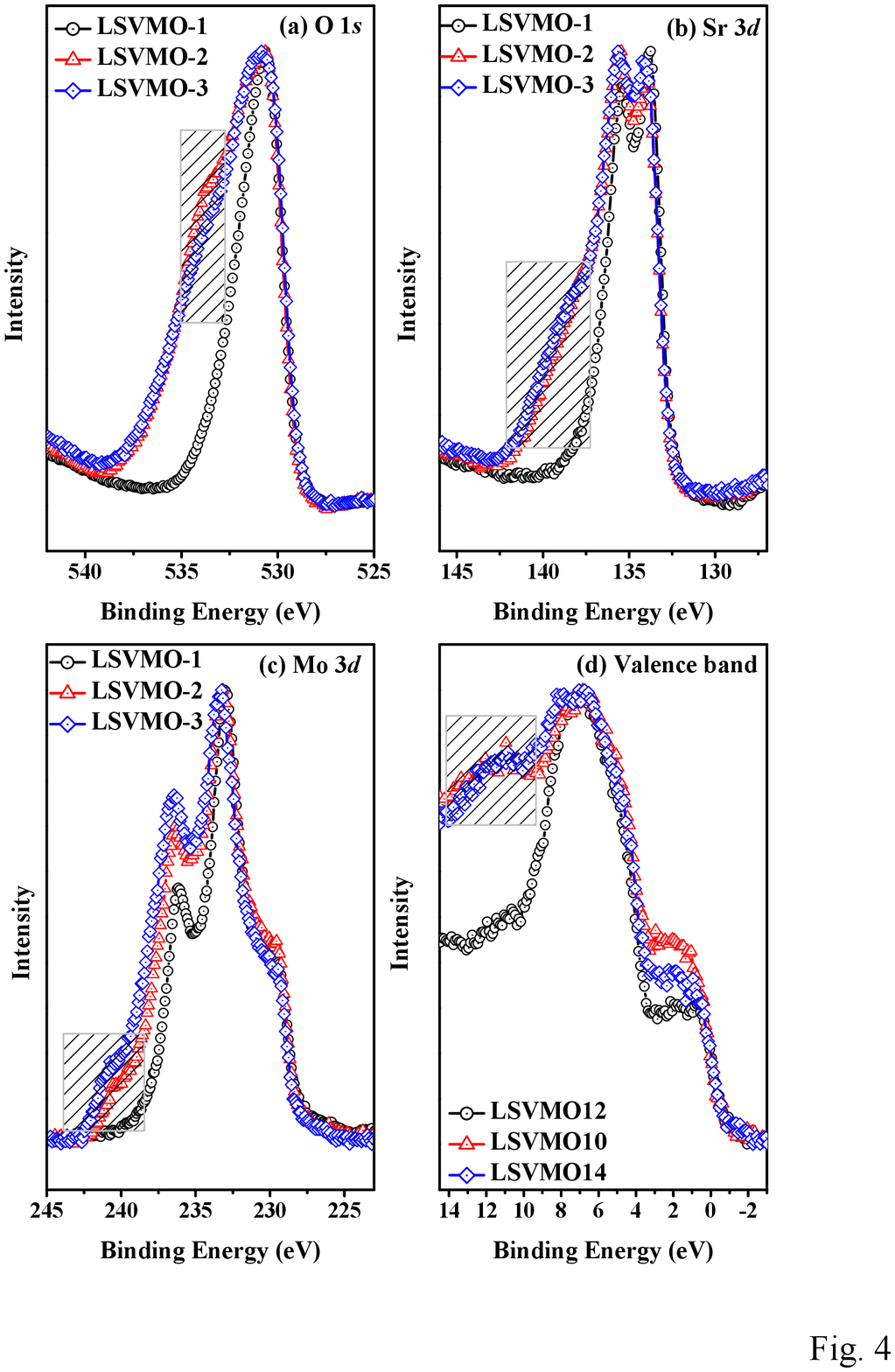}} \\
%\vspace{-1 in}
\caption{(color online) The height normalized O 1$s$ (a), Sr 3$d$ (b), Mo 3$d$ (c) core level and valence band (d) spectra of LSVMO-1 (Open circle), 2 (Up triangle) and 3 (Open rhombus).}
\end{center}
\end{figure}

\begin{figure}
\begin{center}
\centering
\resizebox{10cm}{!}
{\includegraphics*[96pt,75pt][530pt,730pt]{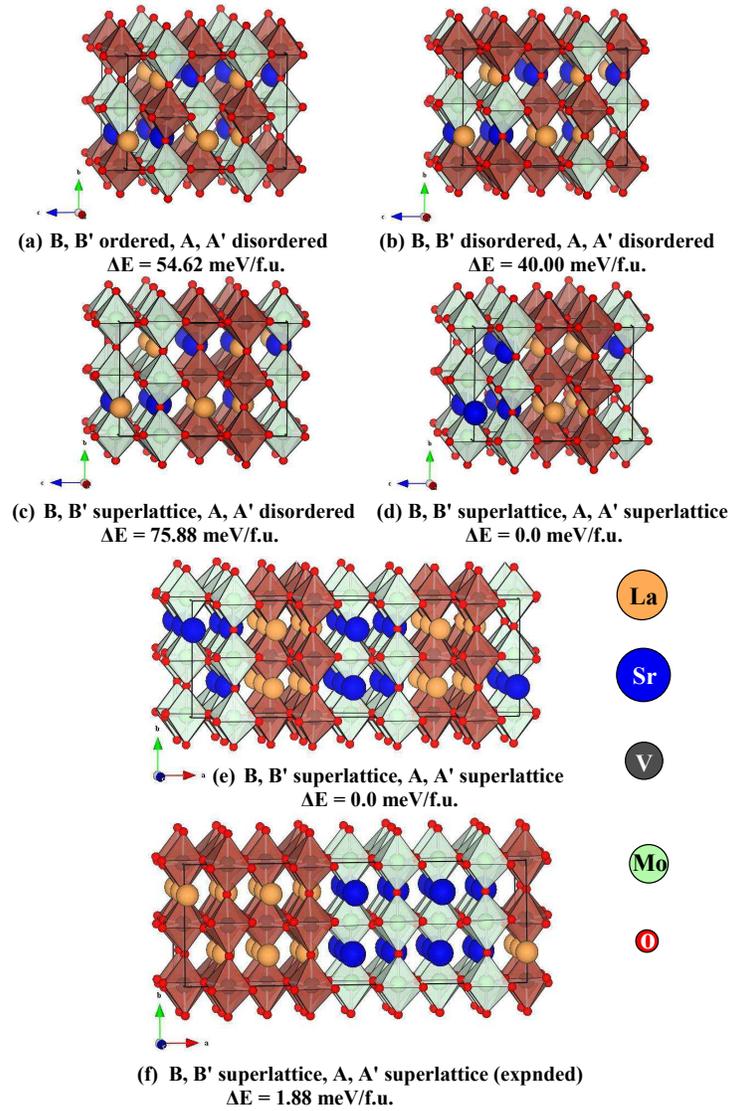}} \\
%\vspace{-1 in}
\caption{(color online) (a)-(f) Different possible structural models of LaSrVMoO$_6$. The corresponding relative stabilization energies ($\Delta$ E) are also shown in the figure.}
\end{center}
\end{figure}

\begin{figure}
\begin{center}
\centering
\resizebox{10cm}{!}
{\includegraphics*[1pt,1pt][318pt,451pt]{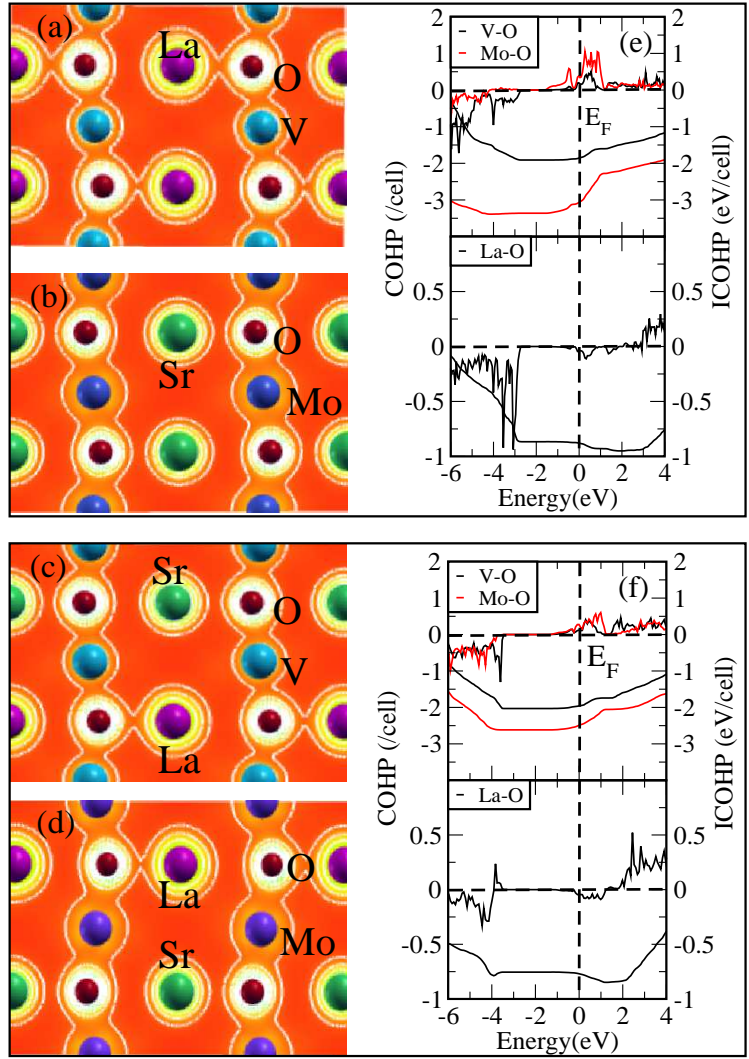}} \\
%\vspace{-1 in}
\caption{(color online) Panels (a,b) show charge density plots for structure (d), while (c,d) show charge density plots for structure (c). Panels (e,f) show off-site COHPs and integrated COHPs (ICOHP) per bond for the nearest neighbors Mo-O, V-O and La-O for structures (d) and (c), respectively. All energies are measured with respect to the Fermi energy.}
\end{center}
\end{figure}

\end{document}